
\input amstex
\input vanilla.sty
\magnification=1200
\TagsOnRight
\def\wh{\widehat}

\def\D{\Cal{D}}
\def\C{\Cal{C}}
\def\w{\text{w}}

\def\ore{\overrightarrow{=}}
\def\ole{\overleftarrow{=}}
\def\om{\omega}

\def\noi{\noindent}

\line{\hfil SB/F/192-92}
\vskip 1cm
\centerline{\bf MASSIVE VECTOR CHERN-SIMONS GRAVITY}
\vskip 2cm
\centerline{\it C. Aragone and P. J. Arias}
\centerline{\it Departamento de F\'{\i}sica, Universidad Sim\'on Bol\'{\i}var}
\centerline{\it Apartado 89000 Caracas 10800-A}
\centerline{\it and}
\centerline{\it A. Khoudeir}
\centerline{\it Laboratorio de F\'{\i}sica Te\'orica,
Departamento de F\'{\i}sica}
\centerline{\it Universidad de los Andes, Apartado 5100, M\'erida}
\vskip 2cm
\centerline{\bf ABSTRACT}
\vskip .3cm
We present a second order gravity action which consists of ordinary Einstein
action augmented by a first-order, vector like, Chern-Simons quasi
topological term. This theory is ghost-free and propagates a pure spin-2 mode.
It is diffeomorphism invariant, although its local Lorentz invariance has
been spontaneuosly broken.

\newpage

In three dimensions, it has been pointed out by Deser, Jakiw and
Templeton$^{[1]}$ that addition of the tensorial topological Chern-Simons
term $S_{TCS}$ $\sim <\om\partial\om +\om^3>$ to the Einstein action $S_E$
yields a gauge invariant, ghost free, pure spin-2 massive theory. In this
paper we present a softer possibility, which also gives rise to a massive
spin-2 theory. Instead of the tensorial $CS$ term we introduce the vectorial
$CS$ term constructed out of the dreibein variables $e^a=dx^re_r^a$
$$
S_{VCS}\ore\mu(2\kappa^2)^{-1}<e_p^a\epsilon^{prs}\partial_re_{sa}>.\tag 1
$$

Here $\mu$ is the topological mass of the system, $\kappa$ is the 3-d
gravitational costant, $e_{sa}=e_s{}^b\eta_{ba}$, $\eta_{ba}$ is the flat
Lorentz metric ($-$++) and $\epsilon^{prs}$ is the Levi-Civita density,
$\epsilon^{012}=+1$.

This action is diffeomorphism invariant and it is not local Lorentz invariant.
It is topological in its world indices. It can not be regarded as fully
topological because it needs the flat metric $\eta^{ab}$ to be a good
invariant. Neither it is locally conformally invariant. As it happens with the
other two actions we mentioned before, $S_{VCS}$ alone does not contain local
excitations. The full action we postulate here is
$$
S\ore (2\kappa^2)^{-1}<e_{pa}\epsilon^{pmn}R_{mn}{}^a(\om )>+S_{VCS}\ole
S_E+S_{VCS}\tag 2
$$
where $R_{mn}{}^a\ore \partial_m\om_n{}^a-$
$\partial_n\om_m{}^a-\epsilon^a{}_{bc}\om_m{}^b\om_n{}^c$ is the planar
Riemann tensor, $\om^a\ore dx^r\om_r{}^a$ and  $e^a\ore$ $dx^re_r{}^a$ being
respectively the affinity and the dreibein  one-forms.

In spite that Einstein action $S_E$ is both local-Lorentz and diffeomorphism
invariant, and the vector-CS term is only diffeomorphism invariant, complexive
action $S$ is just diffeomorphism invariant too.

The situation is similar with massive tensor CS-gravity, which is the sum of
$S_{TCS}-S_E$. This system is not locally conformal invariant due to the
non conformal invariance of the Einstein action. There is a hierarchy of the
local symmetries, starting with the tensorial $CS$ term
$$
S_{TCS}\ore (2\mu\kappa^2)^{-1}<\om_{pa}\epsilon^{pmn}\partial_m
\om_n{}^a-3^{-1}\epsilon^{pmn}\epsilon_{abc}\om_p^a\om_m^b\om_n^c>\tag 3
$$
which is locally conformal, Lorentz, and diffeomorphism invariant. Then it
comes ordinary Einstein action (2) locally Lorentz and diffeomorphism
invariant and finally one has $S_{VCS}$ which is only diffeomorphism
invariant.

Each of these actions alone has non local excitations. However massive
tensorial CS gravity has a pure spin-2 content. Massive vectorial CS gravity,
eq. (2), will be shown to have a pure spin-2 content too. One might even go a
step further and lose all gauge invariances.

In that case, one ends up with self-dual gravity$^{[2]}$, a first order action
on flat three dimensional Minkowski space having no gauge invariance, and
a ghost-free, pure spin-2 content.

Independent variations of $\om_p{}^a$, $e_p{}^a$ in $S$ yield the standard
torsionless value of $\om_p{}^a$ in terms of the dreibein variables.
$$
{}_3e\om_p^{}a=e_{pb}e_q{}^a\epsilon^{qrs}\partial_re_s{}^b-
2^{-1}e_p{}^ae_{qb}\epsilon^{qrs}\partial_re_s{}^b\tag 4
$$
and the (second order in $e_p{}^a$) field equations
$$
E^{pa}\ore \epsilon^{pmn}R_{mn}{}^a(\om )+2\mu \epsilon^{pmn}\partial_m
e_n{}^a =0.\tag 5
$$
Their associated Bianchi identities read
$$
\partial_pE^{pa}-\epsilon_{bc}{}^a\om_p{}^bE^{pc}+
2\mu \epsilon^{nrs}\om_n{}^a\om_r{}^be_{sb}=0\tag 6
$$

Insertion of $\om_p{}^a$ as given by eq. (4) into eq. (5) leads to the second
order field equations which determine the dynamics of the system. The Riemann
tensor has now a source $\sim \mu\epsilon^{pmn}\partial_m e_n{}^a$ which makes
it locally non trivial.

Consequently now we have the possibility of local excitations, as we will show
below. Physical variations of the dreibein variables under small
diffeomorphisms are $\delta e_r{}^a\sim D_r\xi^a$.

In order to understand the physical content of this theory it is convenient to
analyze the associated linearized system, which can be obtained in
straighfoward manner by intoducing $e_{pa}=\eta_{pa}+\kappa h_{pa}$,
$\om_p{}^a=\kappa \om_p{}^a$ in action (2). $S$ then becomes
$$
2S^{Lin}=<2\om_p{}^a\epsilon^{pmn}\partial_mh_{na}-\eta_{pa}\epsilon^{pmn}
\epsilon^a{}_{bc}\om_m{}^b\om_n{}^c>+\mu <h_p{}^a\epsilon^{prs}
\partial_rh_{sa}>\tag 7
$$

We perform a 2+1 decomposition introducing $q_j\ore h_{j0}$. $S^{Lin}$ can
then be written as
$$
\align
2S^{Lin}=<& 2h_{00}\{ \epsilon^{ij}\partial_i\om_j{}^0-\mu\epsilon^{ij}
\partial_iq_j\}+2h_{0l}\{ \epsilon^{ij}\partial_i\om_{jl}+\mu\epsilon^{ij}
\partial_ih_{jl} \}+\\
& +\mu q_i\epsilon^{ij}\partial_0q_j-2\om_i{}^0\epsilon^{ij}\partial_0q_j-
[2\om_{ij}+\mu h_{ij}]\epsilon^{il}\partial_0h_{lj}>+\\
&+<\om_{jj}\om_{ll}-\om_{ij}\om_{ji}>+\\
&+<2\om_0{}^0[\epsilon^{ij}\partial_iq_j+\om_{jj}]+
2\om_0{}^l[\epsilon^{ij}\partial_ih_{jl}-\om_l{}^0]>.\tag 8
\endalign
$$
$\om_0{}^0$, $\om_0{}^l$ constitute multipliers associated with the algebraic
constraints
$$\align
\D_0 & \ore \epsilon^{ij}\partial_iq_j+\om_{jj}=0,\tag 9a\\
\D_l & \ore \epsilon^{ij}\partial_ih_{jl}-\om_l{}^0=0.\tag 9b
\endalign
$$

They provide the respective values of $\om_{jj}$, $\om_l{}^0$ in terms of the
coordinates $h_{jl}$, $q_i$. Then we observe that $h_{00}$, $h_{0l}$ are
Lagrange multipliers too. They are asocciated with the diferential constraints
$$\align
\C^0 & \ore \epsilon_{ij}\partial_i\om_j{}^0-\mu\epsilon_{ij}\partial_iq_j=0,
\tag 10a\\
\C^l & \ore \epsilon_{ij}\partial_i\om_{jl}+\mu\epsilon^{ij}\partial_ih_{jl}=0.
\tag 10b
\endalign
$$

To achieve the unconstrained formulation we introduce the $T+L\ $
2-dimensional decompositions
$$\align
q_j = &(i\wh{\partial})_jq^T+\wh{\partial}_jq^L,\tag 11a\\
h_{ij}=&(i\wh{\partial})_ih^T_j+\wh{\partial}_ih^L_j\\
\ore & (i\wh{\partial})_i(i\wh{\partial})_jh^{TT}+
(i\wh{\partial})_i\wh{\partial}_jh^{TL}+\\
+&\wh{\partial}_i(i\wh{\partial})_jh^{LT}+\wh{\partial}_i\wh{\partial}_j
h^{LL}\tag 11b
\endalign
$$
and similarly for $\om_{ij}$, where $\wh{\partial}_i$ is the unit gradient,
$\wh{\partial}_j\ore \rho^{-1}\partial_j$, $\rho\ore (-\Delta_2)^{1\over{2}}$,
$(i\wh{\partial})_j\ore$ $-\epsilon_{jl}\wh{\partial}_l$.

Eq. (9b) gives the value of $\om_l{}^0$,
$$
\om_l{}^0=-\rho h^T{}_l.\tag 12
$$

Inserting this expression for $\om_l{}^0$ into $\C^0$, eq. (10a) provides
$q^T$
$$
q^T=-\mu^{-1}\rho h^{TT}.\tag 13
$$

The vectorial differential constraint $\C^l$ determines $\om^T_l$
$$
\om^T{}_l=-\mu h^T{}_l.\tag 14
$$

Finally, getting back to $\D_0$ and taking into account eqs. (13) (14) we
obtain $\om^{LL}$,
$$
\om^{LL}=(1+\mu^{-2}\rho^2)\mu h^{TT}.\tag 15
$$

Introducing all this information into $S^{Lin}$ one is led to its (almost
canonical) unconstrained expression

$$
2S^{Lin}=<(\mu h^{TL}-\om^{LT})2\dot{h}^{TT}-2h^{TT}(\mu^2+\rho^2)h^{TT}+
2\mu h^{TL}\om^{LT}>.\tag 16
$$

\vskip .2cm
Redefining $2h^{TT}\to h^{TT}$ and introducing the canonical momenta
$p\ore \mu h^{TL}-\om^{LT}$ eq. (16) transforms into

$$
2S^{Lin}=<p\dot{h}^{TT}-2^{-1}h^{TT}(\mu^2+\rho^2)h^{TT}+
2\om^{LT}(p+\om^{LT})>.\tag 17
$$

\vskip .2cm
\noi
which, after making independent variations of $\om^{LT}$ attains the
standard canonical structure for the unique massive excitation carried out by
$h^{TT}$.

Note that $p\sim \mu h^{TL}$. The self-dual character of the action due to the
presence of the vectorial $CS$ term constrains the tranverse part of $h_{ij}$,
$h^T_j=(i\wh{\partial}_j)h^{TT}+\wh{\partial}_jh^{TL}$, to carry on both the
local physical excitation $h^{TT}$ and its canonical momenta. Moreover action
(17) does not contain either of the longitudinal gauge sensitive variables
$h^{LL}$, $h^{LT}$, $q^L$ (as it must happen) because of the gauge invariance
of $S^{Lin}$ with respect to $\delta h_{ij}=\partial_i\xi_j$.

Now we focus on the curved action $S$. It is convenient to introduce the 2+1
variables $e_{i\bar{j}}$, $q_i\ore e_{i\bar{0}}$, $e^0{}_{\bar{0}}=n^{-1}$,
$e_{0\bar{j}}=-\nu_{\bar{j}}$ and the two dimensional inverse of
$e_{i\bar{j}}$, $\ e^j{}_{\bar{l}}$:
$\ e^j{}_{\bar{l}}e_j{}^{\bar{k}}=\delta_{\bar{l}}^{\bar{k}}$.

It is inmediate to realize that ${}_3e=\epsilon^{rst}e_r{}^{\bar{0}}
e_s{}^{\bar{1}}e_t{}^{\bar{2}}={}_2en$,
$e_0{}^{\bar{0}}=n+\nu_{\bar{j}}e^{i\bar{j}}p_i$. In terms of these variables
$S$ has the form
$$\align
2\kappa^2S=< & 2\om_0{}^{\bar{0}}\{ \epsilon^{ij}\partial_iq_j+
{}_2ee^{i\bar{j}}\om_{i\bar{j}}\}\\
&+2\om_0{}^{\bar{l}}\{ \epsilon^{ij}\partial_ie_{j\bar{l}}+{}_2eq_j
(e^{j\bar{l}}\om_2-\om^{\bar{l}j})-{}_2ee^h{}_{\bar{l}}\om_h{}^{\bar{0}}\}\\
&-2e_0{}^{\bar{0}}\{ \epsilon^{ij}\partial_i\om_j{}^{\bar{0}}-\mu\epsilon^{ij}
\partial_iq_j+2^{-1}{}_2e\om^{\bar{l}j}\om_{j\bar{l}}-
2^{-1}{}_2e\om^{\bar{l}}{}_{\bar{l}}\om^{\bar{j}}{}_{\bar{j}}\}\\
&+2e_0{}^{\bar{l}}\{ \epsilon^{ij}\partial_i\om_{j\bar{l}}+
\mu\epsilon^{ij}\partial_ie_{j\bar{l}}+{}_2e
(\om^{\bar{l}j}-e^{j\bar{l}}\om )\om_j{}^{\bar{0}}\}\\
&+[\mu q_i-2\om_i{}^{\bar{0}}]\epsilon^{ij}\partial_iq_j-
[2\om_{i\bar{l}}+\mu e_{i\bar{l}}]\epsilon^{ih}\partial_0e_{h\bar{l}}>\\
=< & 2\om_0{}^{\bar{0}}\D_{\bar{0}}+2\om_0{}^{\bar{l}}\D_{\bar{l}}+
2e_{0\bar{0}}\C^{\bar{0}}+2e_{0\bar{l}}\C_{\bar{l}}+\\
&+[\mu q_i-2\om_i{}^{\bar{0}}]\epsilon^{ij}\partial_0q_j-
[2\om_{i\bar{l}}+\mu e_{i\bar{l}}]\epsilon^{ij}\partial_0e_{j\bar{l}}>.\tag 18
\endalign
$$

The interacting structure makes the constraints to become highly non linear,
especially the differential ones $\C^{\bar{0}}$, $\C_{\bar{l}}$. The dynamical
germ however remains stable, keeping its quadratic self-dual structure
identical to the corresponding terms shown in eq. (7). The curved algebraic
constraints are
$$\align
\D_{\bar{0}}\ore &
\epsilon^{ij}\partial_iq_j+{}_2ee^{i\bar{j}}\om_{i\bar{j}}=0,
\tag 19\\
\D_{\bar{l}}\ore &\epsilon^{ij}\partial_ie_{j\bar{l}}+{}_2eq_j
(e^{j\bar{l}}\om_2-\om^{\bar{l}j})-{}_2ee^h{}_{\bar{l}}\om_h{}^{\bar{0}}=0;
\tag 20
\endalign
$$
while the differential ones (stemming in the Bianchi identities eqs. (6))
take the aspect

$$\align
E^{0\bar{0}}=\C^{\bar{0}}\ore & \epsilon^{ij}\partial_i\om_j{}^{\bar{0}}-
\mu\epsilon^{ij}\partial_iq_j+2^{-1}{}_2e\om^{\bar{l}j}\om_{j\bar{l}}-
2^{-1}{}_2e\om_2^2=0,\tag 21\\
E^{0\bar{l}}=\C_{\bar{l}}\ore & \epsilon^{ij}\partial_i\om_{j\bar{l}}+
\mu\epsilon^{ij}\partial_ie_{j\bar{l}}+{}_2e(\om^{\bar{l}j}-
e^{j\bar{l}}\om_2)\om_j{}^{\bar{0}}=0.\tag 22
\endalign
$$

\vskip .2cm
To understand the dynamics in the curved case we choose the transverse gauge
for both $e_{ij}$ and $q_i$. This means taking $q^L=h^{LT}=h^{LL}=0$ when we
decompose $q_i$, $e_{ij}$ acoording to eqs. (11), i.e.
$$\align
q_j &=(i\wh{\partial})_jq^T,\tag 23a\\
e_{i\bar{j}}&\ore (i\wh{\partial})_ih^T{}_{\bar{j}}\ore
(i\wh{\partial})_i(i\wh{\partial})_jh^{TT}+(i\wh{\partial})_i\wh{\partial}_j
h^{TL}.\tag 23b
\endalign
$$

Moreover:
$$
\om_i{}^{\bar{0}}=(i\wh{\partial})_i\w^T+\wh{\partial}_i\w^L.\tag 24
$$

The dynamical germ becomes
$$\align
<[\mu q_i-2\om_i{}^{\bar{0}}]&\epsilon^{ij}\partial_0q_j-
[2\om_{i\bar{l}}+\mu e_{i\bar{l}}]\epsilon^{ij}\partial_0e_{j\bar{l}}>=\\
=& <-2\w^L\dot{q}^T-2\om^{LL}\dot{h}^{TL}-2\om^{LT}\dot{h}^{TT}>,\tag 25
\endalign
$$
while the constraints (18)(19)(20)(21) acquire the form:
$$\align
\D_{\bar{0}}\sim &\rho q^T={}_2e\om_2,\tag 26\\
&\\
\D_j\sim &\om_j{}^{\bar{0}}=-e^{-1}(\rho h^T{}_{\bar{l}})e_{j\bar{l}}+
(i\wh{\partial})_lq^T\cdot (\delta^l{}_j\om_2-\om_j{}^l),\tag 27\\
&\\
\C^{\bar{0}}\sim &-\rho \w^T+\mu\rho q^T+2^{-1}{}_2e\om^{\bar{l}j}
\om_{j\bar{l}}-2^{-1}{}_2e\om^2_2=0,\tag 28\\
&\\
\C_{\bar{l}}\sim &-\rho\om^T{}_{\bar{l}}-\mu\rho h^T{}_{\bar{l}}+
{}_2e(\om^{\bar{l}j}-e^{j\bar{l}}\om_2)\om_j{}^{\bar{0}}=0.\tag 29
\endalign
$$

Observe that $e_{j\bar{l}}$ is given in terms of $h^{TT}$, $h^{TL}$ as it is
shown
by eq. (23b). Consequently eq. (27) provides the value of $\om_j{}^{\bar{0}}$
in terms of $h^{TT}$, $h^{TL}$, $q^T$, $\om^{TT}$, $\om^{TL}$, $\om^{LT}$,
$\om^{LL}$. This is the role played by $\D_j$. We then go to
$\C^{\bar{0}}$, eq. (27) which is regarded as an equation to solve for
$q^T=q^T(h^{TT},h^{TL},\om^{TT},\om^{TL},\om^{LT},\om^{LL})$. Once we obtain
$q^T$, it is introduced back into $\D_j$ (27) which then yields
$\om_j{}^{\bar{0}}=\om_j{}^{\bar{0}}
(h^{TT},h^{TL},\om^{TT},\om^{TL},\om^{LT},\om^{LL})$. This functional value of
$\om_j{}^{\bar{0}}$ is introduced back into $\C_{\bar{l}}$ (eq. (29)). From
them we solve $\om^{TT}$, $\om^{TL}$ in terms of the remaining four
variables $h^{TT}$, $h^{TL}$, $\om^{LT}$, $\om^{LL}$. These values can also be
substituted into the previous functionals $q^T$ and $\om_j{}^{\bar{0}}$ so we
have  $\om_j{}^{\bar{0}}$, $q^T$, $\om^{TT}$, $\om^{TL}$ expressed as
functionals of  ($h^{TT}$, $h^{TL}$, $\om^{LT}$, $\om^{LL}$).
If we then insert all these expressions into $\D_{\bar{0}}$, we arrive to a
functional equation which determines $\om^{LL}=\om^{LL}$
($h^{TT}$, $h^{TL}$, $\om^{LT}$).
And back again to the previous expressions we will get
$\om_j{}^{\bar{0}}$, $q^T$, $\om^{TT}$, $\om^{TL}$, $\om^{LL}$ as functionals
of
($h^{TT}$, $h^{TL}$, $\om^{LT}$).

After all this procedure is done we will have the unconstrained action in
terms of the physical excitation $h^{TT}$, the canonical momenta
$p\sim {\mu h^{TL}-\om^{LT}}$ and an auxiliary variable $\om^{LT}$.
Similarly to what we have explicitly seen in the linearized case, we conjeture
that its field equation will be a constraint which can ve solved for $\om^{LT}$
in terms of $h^{TT}$ and the canonical momenta $p$, giving rise to the final,
unconstrained canonical action. This point deserves a more detailed analysis.

Summing up, we have presented a second order, diffeomorphism invariant
action containing a first order CS-term which contains one local degree of
freedom corresponding to a propagating spin-2 massive excitation. There is a
substancial difference between vector Chern-Simons gravity and topological
massive gravity arising from the fact that here we have the Einstein action
with the standard sign whereas in topological massive gravity the Einstein's
action must be written with the opposite sign$^{[1][3]}$.

\newpage

\centerline{\bf REFERENCES}
\vskip .3cm

\item{[1]}S. Deser, R. Jackiw and S. Templeton, Ann. of Phys {\bf 140} (1982)
372, (E) {\bf 185} (1988) 406.
\item{[2]}C. Aragone and A. Khoudeir, Phys. Lett. {\bf B173} (1986) 141;
{\it Quantum Mechanics of Fundamentals Systems 1}, ed. C. Teitelboim, Plenum
Press, New York (1988) pp17.
\item{[3]}C. Aragone, Class Q Gravity {\bf 4} (1987) L1.
\end